\documentclass[5p]{elsarticle}

\usepackage{hyperref, lineno}
\modulolinenumbers[5]

\journal{Astronomy and Computing}

\biboptions{authoryear}

\begin{document}

\begin{frontmatter}

\title{PHOTOMETRYPIPELINE: An Automated Pipeline for Calibrated Photometry}

\author{Michael Mommert}
\address{Northern Arizona University, Department of Physics and Astronomy, Flagstaff, AZ 86011, USA}
\ead{michael.mommert@nau.edu}




\begin{abstract} PHOTOMETRYPIPELINE (PP) is an automated pipeline 
    that produces calibrated photometry from imaging data through
  image registration, aperture photometry, photometric calibration,
  and target identification with only minimal human interaction. 
    PP utilizes the widely used Source Extractor software for source
    identification and aperture photometry; SCAMP is used for image
    registration.  Both image registration and  photometric
  calibration are based on matching field stars with star
    catalogs, requiring catalog coverage of the respective field. A
    number of different astrometric and photometric catalogs can be
    queried online. Relying on a sufficient number of background
  stars for image registration and photometric calibration, PP is
  well-suited to analyze data from small to medium-sized
  telescopes. Calibrated magnitudes obtained by PP are typically
  accurate within
  ${\leq}$ 0.03 mag and astrometric accuracies are of the order of
  0.3~arcsec relative to the catalogs used in the registration. The
  pipeline consists of an open-source software suite written in Python
  2.7, can be run on Unix-based systems on a simple desktop machine,
  and is capable of realtime data analysis. PP has been developed for
  observations of moving targets, but can be used for analyzing point
  source observations of any kind.
\end{abstract}


\end{frontmatter}


\section{Introduction} \label{sec:intro}

Telescopes across the globe acquire massive amounts of imaging data
every night. While the underlying science goals vary widely in these
observations --- from deep observations of extragalactic targets to
short observations of rapidly spinning asteroids --- the
immediate objective of most observations is similar: obtaining
reliable and calibrated brightness measurements of usually faint
point sources. This objective requires not only good seeing and
transparency conditions, as well as more or less extensive planning in
order to address the science goal in the most efficient way, but also
a sophisticated and accurate reduction and analysis of the acquired
data.

Large observatories often provide support in the reduction and
analysis of their data. However, the majority of imaging data have been
--- and still is --- acquired with telescope apertures of a few meters
or smaller. Smaller telescopes are usually easier to access because
they are more numerous, but the observer is often left alone in the
data reduction and analysis process. This factor leads to large
amounts of imaging data from smaller telescopes being left unanalyzed
as their proper analysis is not considered worth the effort, or
because observing conditions were not ideal. The availability of an
automated and robust software pipeline to exploit these data would
simplify access to this data treasure trove.

I present  PHOTOMETRYPIPELINE (PP), a Python-based, open-source
software suite that provides automated and calibrated point-source
photometry of imaging data, specifically designed for small to
medium-sized telescopes. PP provides image registration, photometric
analysis and calibration for both fixed and moving targets with only
minimal user interaction. The pipeline can be run on Unix-based
systems, ranging from desktop machine to larger and more capable
  machines. PP was
originally designed to obtain photometry of asteroids, but can be
applied to observations of any point-sources, including stars,
extragalactic sources, artificial satellites, and space debris. Due to
its modular and flexible design, it can be modified to work with data
from nearly any professional telescope.

PP is available for download on GitHub\footnote{\url{https://github.com/mommermi/photometrypipeline}}. Since PP is
still evolving, refer to the online documentation\footnote{\url{http://mommermi.github.io/pp/index.html}} for
up-to-date information. This document describes the functionality of
PP Version 1.0 as of 30 November 2016.
Also refer to the documentation for installation guides and additional
support.

\section{Methods and Implementation}
\label{lbl:methods}

\subsection{Overview}

\begin{figure}[t]
\centering
\includegraphics[width=\linewidth]{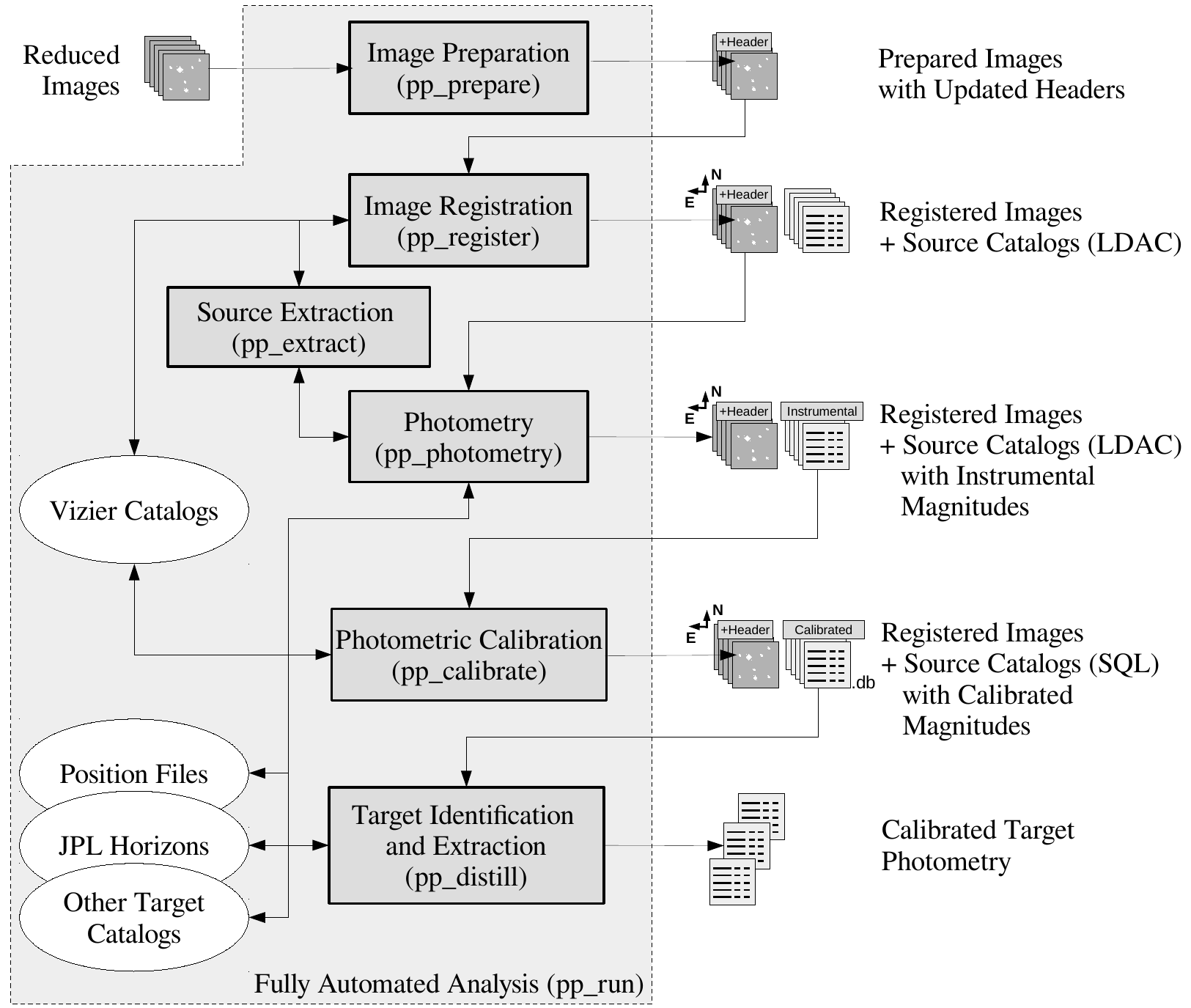}
\caption{PP work flow diagram. The shaded area indicates the sequence
  of tasks run by the automatic analysis routine {\tt pp\_run}; the
  same sequence should be used if the individual tasks are run
  separately. White ovals indicate resources that are required in the
  analysis. The individual pipeline steps are discussed in Section
  \ref{lbl:methods}.\label{fig:flowchart}}
\end{figure}

The pipeline is implemented as a suite of Python 2.7 scripts. It makes
use of Python packages that are freely available and easy to install
through the Python Package
Index\footnote{\url{https://pypi.python.org/pypi}}; required packages
include NumPy\footnote{\url{http://www.numpy.org/}},
SciPy\footnote{\url{https://www.scipy.org/}},
astropy\footnote{\url{http://www.astropy.org/}},
matplotlib\footnote{\url{http://matplotlib.org/}}, and
CALLHORIZONS\footnote{\url{https://github.com/mommermi/callhorizons}
  and Section \ref{lbl:callhorizons}}.  For specific tasks,
open-source software has been integrated into the pipeline. That
software is called within the Python environment, but has to be
installed on the machine prior to running the pipeline. The required
auxiliary software is Source
Extractor\footnote{\url{http://www.astromatic.net/software/sextractor}}
\citep{Bertin1996} and
SCAMP\footnote{\url{http://www.astromatic.net/software/scamp}}
\citep{Bertin2002}, which are introduced below. Both Source
  Extractor and SCAMP are well-tested and widely used within the
  community, providing a robust foundation for the most important
  pipeline features.

PP consists of a number of stand-alone Python scripts that can be
called separately, or can be run automatically. Figure
\ref{fig:flowchart} presents the work flow of the automated pipeline
routine ({\tt pp\_run}). The same sequence should be used if the
individual tasks are called separately in order to meet the respective
file dependencies. {\tt pp\_run} is designed to run automatically on
the majority of all data provided; this implies that it might not run
successfully on non-ideal data (see Section \ref{lbl:limits} for a
discussion). Running the individual pipeline tasks separately, using
fine-tuned parameters differing from the defaults used by {\tt
  pp\_run}, might be necessary to improve the outcome for non-ideal datasets.

The runtime of the pipeline depends on a number of different
parameters, including the memory and computing power of the machine,
the image size, the number of images, the background star density, and
the number and nature of targets in each field. For instance, running
the example data presented in Section \ref{lbl:results} (79
700~pixel$\times$700~pixel images, one target in the field) through the default
pipeline (using {\tt pp\_run}) on a quad-core 1.9~GHz laptop running
Ubuntu Linux 16.04 takes less than 20~minutes. With minor modifications, PP is
able to provide real-time data analysis.

Every step of the image analysis process is thoroughly documented and
summarized in a ``diagnostics'' HTML file that is created on-the-fly. This
webpage allows for inspection of each pipeline process and serves as a
validation of the data quality. Furthermore, it allows for
identification of data affected by background sources, artifacts, and
target mis-identification.

In order to provide the best possible results, PP should be run on
fully reduced image data, which includes flat fielding and bias
correction, as well as trimming of the data. It is mandatory that all
frames coming from the same instrument have the same image dimensions.

\subsection{Pipeline Tasks}

The individual pipeline tasks shown in Figure \ref{fig:flowchart} are
described in detail below. Refer to the documentation
provided online for more details on how to use these tasks.

\subsubsection{Image Data Preparation -- {\tt pp\_prepare}}

In order to allow for the degree of automation provided by PP, it
heavily relies on properly populated FITS image headers. Since every
instrument/telescope combination uses slightly different formats, each
combination has to be set up before data can be run through the
pipeline. This setup consists of tailored parameter files for Source
Extractor and SCAMP (see below), as well as a dictionary that
translates header information into a format that is readable by the
pipeline. In order to exploit its full potential, PP requires
information on the telescope pointing and the date and time of the
observations, as well as the detector pixel scale, detector binning,
the official Minor Planet
  Center\footnote{\url{http://minorplanetcenter.net}} (MPC)
identifier of the observatory, the used photometric filter, and the
target name to be present in each FITS image header.

Pipeline task {\tt pp\_prepare} identifies the used instrument and
then reads, translates, and modifies the required FITS header keywords
into a common format that is independent of the instrument and
readable by all pipeline tasks. It also removes existing plate
solutions in the World Coordinate System \citep[WCS,
see][]{Greisen2002, Calabretta2002} format from the header and
implants a zero-th order solution based on the provided image
coordinates, the detector pixel scale, and the typical image
orientation for the respective telescope/in\-strument combination. This
step is crucial for proper image registration (see Section
\ref{lbl:pp_register}).

This approach grants a high degree of flexibility to the pipeline,
making it applicable to a large range of telescopes and instruments.

\subsubsection{Image Registration -- {\tt pp\_extract} and {\tt
    pp\_register}}
\label{lbl:pp_register}

A plate solution for each input image is found using SCAMP, which
computes astrometric solutions based on coarse WCS information in the
FITS image header (provided by {\tt pp\_prepare}), a catalog of all
sources in the field, and a reference catalog. SCAMP works completely
automatic. Field source catalogs are generated using Source Extractor
in the binary Leiden Data Analysis Center (LDAC) catalog
format. Source Extractor identifies field sources based on their
signal-to-noise ratio (SNR) and a minimum number of connected pixels
that exceed that SNR threshold. For each source, Source Extractor
provides the position in image and -- if available -- WCS coordinates,
different flavors of photometry, descriptive flags, and other
diagnostic parameters. Both Source Extractor and SCAMP are integrated
into PP; the user will not have to interact with either of them
directly.

Source extraction is performed by {\tt pp\_extract}, using Source
Extractor, building a LDAC catalog for each input image. Source
Extractor parameters can be controlled through {\tt pp\_extract}. In order to
improve runtime and to exploit multi-CPU architecture, {\tt
  pp\_extract} uses Python's multi-threading capabilities. Despite the
fact that all pipe\-line results rely on this task, normally it will not
be called by the user directly -- although this is possible, and
recommended in specific cases.

LDAC catalogs are read into {\tt pp\_register}, which calls SCAMP to
find and imprint an astrometric solution into each image file.  SCAMP
matches LDAC catalogs against catalogs with astrometric solutions that
it retrieves from the VizieR
service\footnote{\url{http://vizier.u-strasbg.fr/viz-bin/VizieR}} at
the Centre de Donn\'{e}es astronomi\-ques de
Strasbourg\footnote{\url{http://cds.u-strasbg.fr/}}. A large number of
potential catalogs is available (refer to the SCAMP manual for an
overview); PP usually utilizes URAT-1 \citep{Zacharias2015}, 2MASS
\citep{Skrutskie2006}, or USNO-B1 \citep{Monet2003}, providing robust
astrometry. {\tt pp\_register} will try each of the aforementioned
catalogs (in that order) until all input images have been registered
successfully. For each catalog, SCAMP is called twice using the same
catalog, unless all input images have been registered successfully in
the first run; the second run uses additional information created in
the first SCAMP run. Registration fails if the number of field stars
is too small (fewer than 10-20 stars), the images suffer from severe
artifacts, or the initial WCS header solution is too far off
(typically more than ${\sim}$10 arcmin). Typically, image registration
is successful even if the image orientation is previously unknown,
given that the image center position and pixel scale are reasonably
well known. The derived astrometric solution includes image
  distortion corrections as derived by SCAMP; the order of the
  correction terms varies for different telescopes.  The results of
{\tt pp\_register}, including information on each input image and a
thumbnail image indicating the match based on the utilized catalog,
are added to the diagnostics webpage. The reliability of the image
registration provided by SCAMP and PP is verified in Section
\ref{lbl:astrometry_verification}.

\subsubsection{Photometry -- {\tt pp\_photometry}}
\label{lbl:pp_photometry}

\begin{figure}[t]
\centering 
\includegraphics[width=\linewidth]{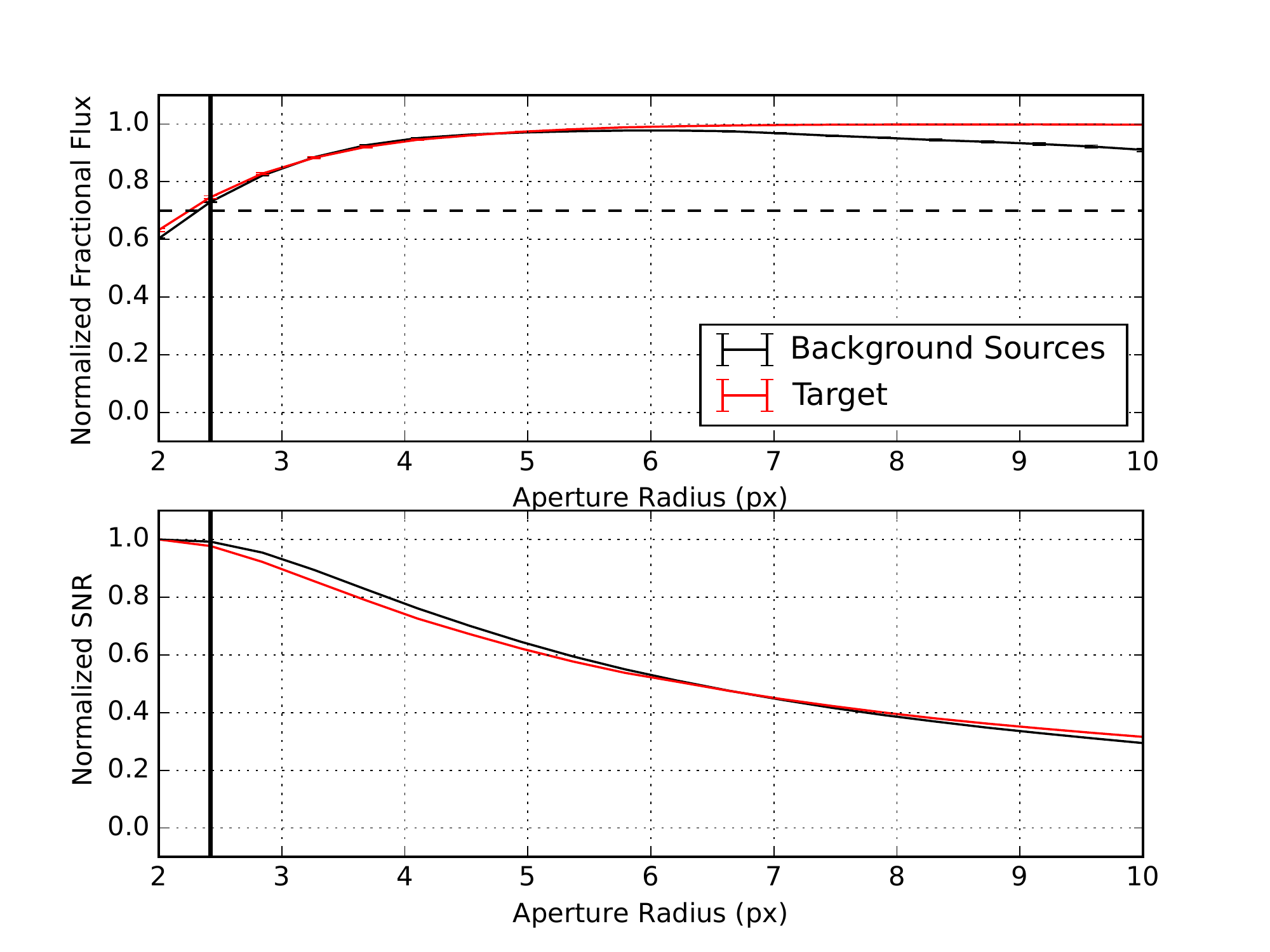}
\caption{Curve-of-Growth diagnostic plot created by PP for one image from
  the data set used in Section \ref{lbl:results}. The top panel shows the normalized fractional flux of the target and the background sources as a function of aperture radius, the bottom panel shows the normalized SNR. The vertical bar indicates the aperture radius that is adopted as optimum aperture radius using the criteria discussed in Section \ref{lbl:pp_photometry}.
  \label{fig:pp_photometry}}
\end{figure}

PP uses aperture photometry performed by Source Extractor. The optimum
aperture radius used in this process is derived in a curve-of-growth
analysis \citep{Howell2000}, in which {\tt pp\_extract} measures the
flux and its corresponding uncertainty for each source in every image
using 20 different aperture radii. The target (or multiple targets) is
identified (see Section \ref{lbl:target_identification}) and its
fluxes are isolated from those of the background sources. The fluxes
for the target and the background sources are averaged separately over
all images as a function of aperture radius and normalized
separately. As shown in Figure \ref{fig:pp_photometry}, the normalized
flux distribution and the SNR distribution have their maximum at
different aperture radii. By increasing the aperture radius, more flux
from each source is included; at the same time, the increase in
aperture area also increases the noise contribution from the
background. Hence, the flux is maximized at the largest aperture
radius, but the SNR is maximized at a radius of typically 1.5 FWHM
\citep{Howell2000}. In the case of moving target observations, the
target or the background sources might be trailed, leading to
additional difficulties.

PP uses by default the following criteria for the optimum aperture
radius (compare to Figure \ref{fig:pp_photometry}): the smallest
aperture radius at which at least 70\% of each of the total target flux
and the total background flux is included and at the same time the
difference between the normalized target and background flux levels is
smaller than 5\%. These criteria make sure that trailing affects the
target photometry at a level that is within the expected uncertainties
of the photometry results. Alternatively, the user can select an
aperture radius manually, or have the pipeline use that aperture radius that
maximizes the SNR in the target or the background stars. Plots comparable
to Figure \ref{fig:pp_photometry} are generated on-the-fly for each
image set processed by {\tt pp\_photometry} as part of the diagnostic
output of the pipeline.

As a result of {\tt pp\_photometry}, LDAC catalogs for each input
image are created that make use of the optimum (or manually selected)
aperture radius; the catalogs contain instrumental magnitudes.

\subsubsection{Photometric Calibration -- {\tt pp\_calibrate}}
\label{lbl:pp_calibrate}

PP provides photometric calibration of each image using background
stars in the same field.  The advantage of the photometric calibration
using field stars is that data from different telescopes can be
compared directly and that transparency and/or seeing variations
can be compensated for.

In order to obtain calibrated photometry, the offset between the
measured instrumental magnitudes and the respective photometric filter
used in the observations -- the magnitude zeropoint -- has to be
determined.  In PP, this is done by comparing the brightness of field
stars to their catalog brightness. Currently, PP supports the Sloan
Digital Sky Survey Data Release 9 \citep[SDSS-R9,][photometric bands:
u, g, r, i, z]{Ahn2012} and the AAVSO Photometric All-Sky Survey
Release 9 \citep[APASS9,][photometric bands: g, r, i, B,
V]{Henden2016} for optical bands, as well as 2MASS for the near-infrared
bands (photometric bands: J, H, Ks), for calibration purposes.

\begin{figure}[t]
\centering
\includegraphics[width=\linewidth]{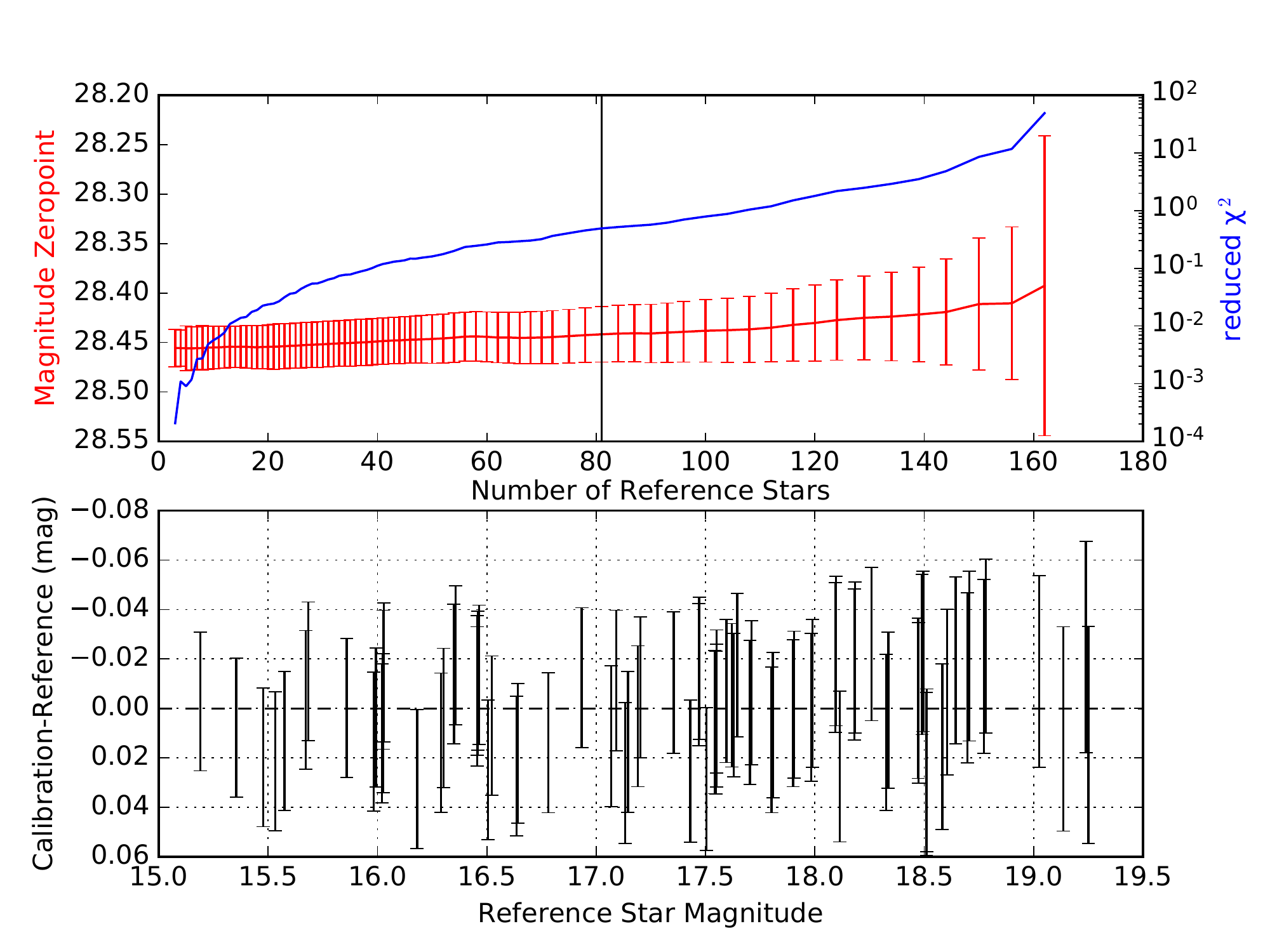}
\caption{Photometric calibration diagnostic plot for one image from
  the data set used in Section \ref{lbl:results}. The top panel shows
  the derived magnitude zeropoint and according reduced $\chi^2$ as a
  function of the number of reference stars used. The vertical line
  indicates the number of reference stars used in the final
  calibration. The bottom panel shows the residuals between the
  measured photometry in the image and the catalog magnitudes for this
  final calibration. Note that the zeropoint uncertainty is limited by
  the residual uncertainties -- the final zeropoint uncertainty is
  0.025~mag in this case. Also note that the zeropoint varies only
  insignificantly as a function of the number of reference
  stars.\label{fig:pp_calibrate}}
\end{figure}

Catalogs are queried from the VizieR service. In the case of optical
data, SDSS is queried first; if there is no SDSS data available, APASS
is used, which covers the majority of the Northern Hemisphere. Catalog
data and field sources are matched based on astrometry; (near-)
saturated or blended sources, as well as sources that do not have
catalog counterparts, no or inaccurate ($\sigma > 0.05$~mag)
photometric data are excluded from the following steps. The magnitude
zeropoint ($m_{zp}$) of each frame is derived as follows (compare to
Figure \ref{fig:pp_calibrate}). The residual $\zeta_i$ between the
catalog magnitude and the instrumental magnitude is calculated for all
$N$ available sources with index $i$. By minimizing

\begin{equation}
\chi^2 = \sum_i^N \frac{(m_{zp}-\zeta_i)^2}{\sigma_{\zeta,i}^2},
\end{equation}

the best-fit $m_{zp}$ is determined, taking into account the
uncertainties of the individual residuals, $\sigma_{\zeta,i}$, which
are root-sum-squares of the uncertainties quoted in the catalogs and
the instrumental uncertainties derived in the aperture
photometry. Since some sources are affected by image artifacts or
blending with background objects, their photometric measurements are
compromised, leading to significantly increased residuals. In order to
account for these outliers, an iterative rejection scheme has been
implemented that removes that source representing the largest outlier
one at a time and recalculates the zeropoint magnitude after each
rejection. The zeropoint magnitude featuring the minimum $\chi^2$ of
all iteration steps and at the same time having $N$ equal at least
50\% of the original number of sources is adopted. The threshold of
50\% is somewhat arbitrary but generally leads to reliable
results. The uncertainty associated with the magnitude zeropoint is
determined as the quadratic sum of the average residual uncertainties
and the weighted standard deviation of the residuals:

\begin{equation}
\sigma_{zp} = \sqrt{  \frac{1}{N} \sum_i^N \sigma_{\zeta, i}^2 + \frac{1}{N}
  \sum_i^N \frac{(\zeta_i - m_{zp})^2}{\sigma_{\zeta,i}^2}}
\end{equation}

Typical zeropoint magnitude uncertainties are of the order of
0.02--0.05~mag, based on typically ${\geq}$10 background sources. At
least three sources are required for a photometric calibration; if no
calibration is possible, instrumental magnitudes are reported.  The
reliability of the photometric calibration is verified in Section
\ref{lbl:photometry_verification}.

Final calibrated photometric measurements for all sources in each
field are written into a queryable
SQLite\footnote{\url{https://sqlite.org/}} database. Plots similar to
Figure \ref{fig:pp_calibrate} are generated for each input image as
part of the diagnostic output.

PP supports transformations between photometric systems. Using
equations provided by \citet{Chonis2008}, SDSS ugriz magnitudes can be
transformed into BVRI magnitudes. Also, using \citet{Hodgkin2009},
2MASS near-infrared bands can be converted into the UKIRT system,
which uses the standard Mauna Kea near-infrared filters. Uncertainties
introduced through these transformations are typically of the order of
a few 0.01~mag (see Section \ref{lbl:photometry_verification} for a
discussion). Additional transformations and photometric catalogs will
be implemented in the future.

Being designed for single-band data, PP does not support a color-term
correction in addition to the derivation of the magnitude
zeropoint. For observatories that automatically obtain data in
different bands, such a correction could be implemented, potentially
improving the overall photometric calibration quality.

\subsubsection{Target Identification and Extraction -- {\tt pp\_distill}}
\label{lbl:target_identification}

Photometry for selected targets is extracted from the SQLite databases using
{\tt pp\_distill}. Targets are identified based on their WCS
coordinates provided through simple text files, or manually provided
fixed coordinates; ephemerides for moving targets are queried from JPL
Horizons using the {\tt CALLHORIZONS} Python module. Other types of
target catalogs, e.g., through a query of SIMBAD\footnote{\url{http://simbad.u-strasbg.fr/simbad/}} or match with other
online resources, will be implemented in the future.

For each target, a photometry file is generated that provides
extensive information, as well as diagnostic output that allows to
inspect the data quality. Furthermore, {\tt pp\_distill} automatically
selects one reasonably bright star as ``control star'', which is
treated the exact same way as any other target and allows for an
assessment of the reliability of the entire analysis procedure.

\subsubsection{Data Products}

The final data products of the pipeline include (1) SQLite database
files with positions, calibrated magnitudes, and additional
information on each source detected for each input frame, (2) ASCII
tables with extracted information on each target that has been
identified, and (3) a summary website with the combined diagnostic
output of each pipeline task.

\section{Example Results}
\label{lbl:results}

In order to test the reliability of the pipeline, PP is run over V
band imaging data of asteroid (2704) Julian Loewe taken with Lowell
Observatory's 42-inch telescope \citep{Oszkiewicz2016} and compared to
results derived with the commercial MPO
Canopus\footnote{\url{http://www.minorplanetobserver.com/MPOSoftware/MPOCanopus.htm}}
software. MPO Canopus also provides astrometric and photometric
calibration of imaging data. In the case of this data set, the
photometric calibration provided by MPO Canopus is based on a
transformation from 2MASS near-infrared magnitudes to optical
magnitudes \citep{Binzel2004} due to a lack of other calibration
stars.

\begin{figure}[t]
\centering
\includegraphics[width=\linewidth]{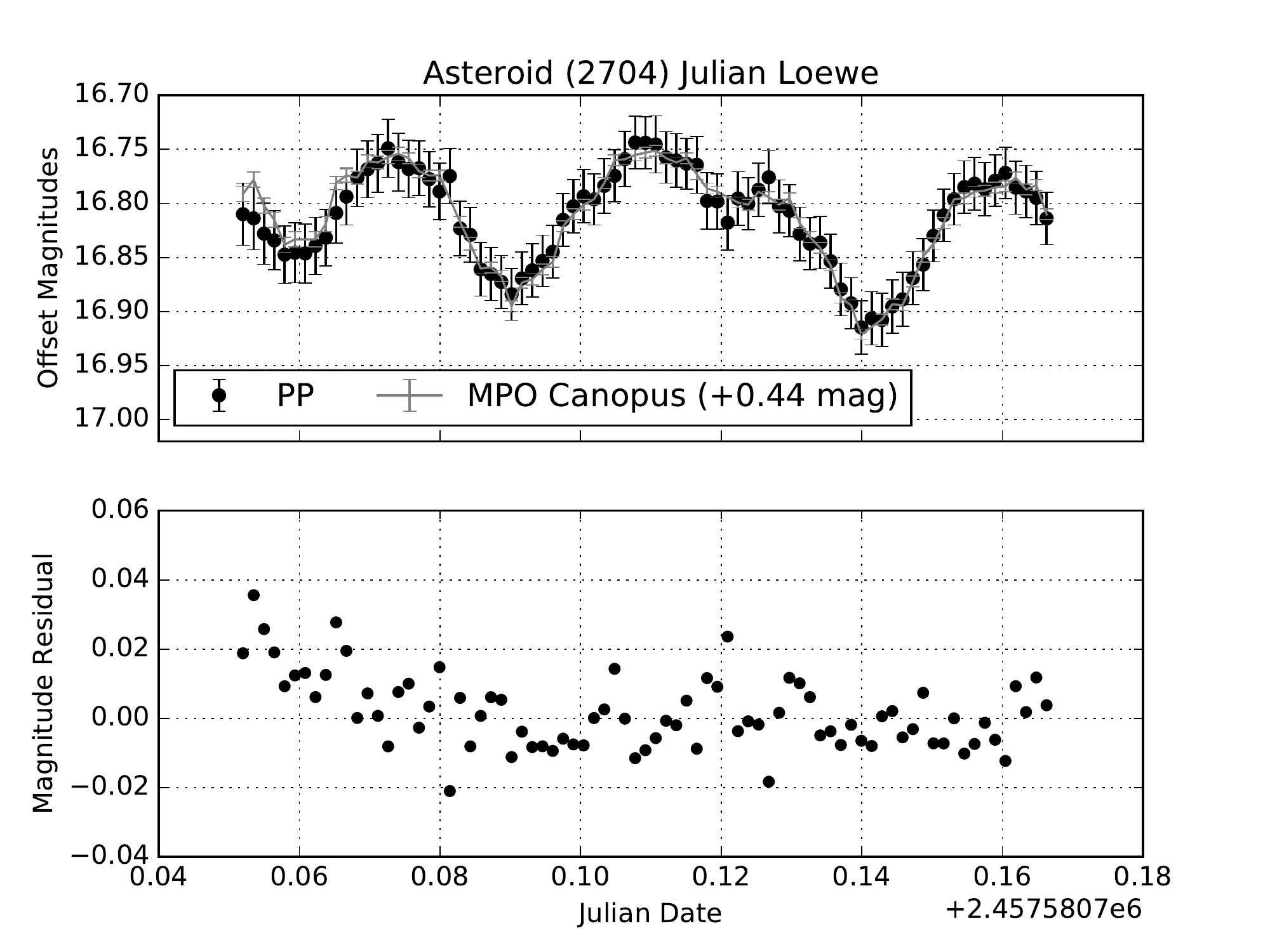}
\caption{Optical (V band) lightcurve of asteroid (2704) Julian Loewe
  as measured by PP and the commercial MPO Canopus software (top
  panel). The relative agreement between both lightcurves is excellent
  (0.01~mag standard deviation of the residuals, lower panel). A
  constant offset between both lightcurve measurements of 0.44~mag is
  found (see text).
\label{fig:results}}
\end{figure}

The resulting lightcurve from PP and the comparison to the MPO Canopus
results are displayed in Figure \ref{fig:results}. The relative
agreement between the lightcurves obtained by the programs is
excellent. The standard deviation of the residuals between both data
sets is of the order of 0.01~mag, which is smaller than the
uncertainties provided by PP and of the same magnitude as the
uncertainties provided by MPO Canopus. Note, however, that there is a
constant offset between the reported magnitudes of 0.44~mag. This
offset is most likely a result of the 2MASS-based photometric
calibration of MPO Canopus: while optical magnitudes are affected by
galactic extinction, this is not the case in the near-infrared bands
provided by 2MASS. Furthermore, this method explicitly assumes
  that all near-infrared magnitudes unambiguously extrapolate to
  optical magnitudes. Hence, a constant offset between the two data
analyses has to be expected.

\section{Discussion}

Section \ref{lbl:results} proves the robustness of PP results in comparison
to other available software. The following sections prove the claimed
photometric and astrometric accuracy of the pipeline.

\subsection{Photometric Accuracy}
\label{lbl:photometry_verification}

The photometric calibration accuracy provided by PP is verified using
observations of standard star fields and their photometry from the
literature. This experiment uses observations of 5 different standard
star fields (centered on 95-43, 95-142, PG0231+051, PG1047+003,
PG1323-086, and RU149) taken with the Discovery Channel Telescope and
its Large Monolithic Imager \citep[LMI,][]{Massey2013} in the V
band. PP is run over the LMI data automatically (using {\tt pp\_run})
and with its default settings; the stars are unambiguously identified
based on finder charts and their positions. The resulting calibrated
photometry, based on the SDSS-R9 and APASS9 catalogs, is compared to
values measured by \citet{Stetson2000}\footnote{\url{www.cadc-ccda.hia-iha.nrc-cnrc.gc.ca/en/community/STETSON/standards}}. Table
\ref{tab:photometry} compiles the mean photometric residuals and the
standard deviations of the residuals measured per field. Since the
instrumental uncertainties for the stars are small (${\leq}0.01$~mag),
the residual statistics reflect the calibration accuracy of the
pipeline.  For all fields, the standard deviation is larger than the
mean residual for both catalogs, suggesting that any systematic
offsets are statistically insignificant.  Both SDSS-R9 and APASS9
provide magnitude zeropoints that are accurate within ${\leq}$0.03~mag
(1$\sigma$ level), using the default pipeline settings. PP
  $1\sigma$ uncertainties are consistently larger than the sample
  standard deviations, proving the conservative nature of the default
  pipeline uncertainties. Johnson--Cousins V
magnitudes are transformed from SDSS g and r magnitudes (see Section
\ref{lbl:pp_calibrate}).

Manual interaction, e.g., decreasing the aperture radius or
reducing/increasing the number of stars used in the photometric
calibration procedure can improve the overall photometric
accuracy. Also, the future availability of high quality photometric
catalogs, e.g., provided by Pan-STARRS, GAIA, LSST, will further
improve the accuracy of the photometric calibration.

\begin{table}
\scriptsize
\begin{tabular}{|rccc|}
\hline \\[-5pt]
{\bf Field} & $\mathbf{N}$ & {\bf SDSS-R9 Residuals} & {\bf APASS9 Residuals} \\
\hline\hline \\[-5pt]
RU149       &   399 &  0.019$\pm$0.024$^{\star}$ (0.039) & -0.014$\pm$0.023 (0.036) \\
PG0231+051  &   20 &  0.002$\pm$0.018 (0.025)  & -0.005$\pm$0.014 (0.036) \\
PG1047+003  &   24 &  0.019$\pm$0.025 (0.032)  & -0.009$\pm$0.025 (0.050) \\
PG1323-086      &   18 &     --- & 0.020$\pm$0.027 (0.065) \\
    95-142  &   30 &  0.013$\pm$0.025 (0.025)  &  0.001$\pm$0.025 (0.043)\\
     95-43  &   27 &  0.013$\pm$0.021 (0.033)  & -0.001$\pm$0.023 (0.035) \\
\hline
\end{tabular}
\caption{Verification of the photometric accuracy. $N$ is the number of stars in the respective field, the residual columns list the mean and the standard deviation of the residuals between the values provided by P.\ Stetson and the measured magnitudes based on the respective catalog with PP; numbers in brackets are the default 1$\sigma$ uncertainties on the magnitude
  zeropoint as derived by PP. Residual means and standard deviations
  are better than 0.03~mag, reflecting the overall calibration
  accuracy that can be achieved with the pipeline using available
  catalogs. ($^{\star}$: SDSS-R9 photometry compromised due to nonuniform
  coverage of SDSS-R9 stars throughout the field.)}
\label{tab:photometry}
\end{table}

\subsection{Astrometric Accuracy}
\label{lbl:astrometry_verification}

Proper astrometric calibration is necessary for unambiguous target
identification, but also provides crucial orbital information for
asteroids and positional information for other targets. PP relies on
SCAMP and available catalogs to establish plate solutions and
astrometric calibration; currently available catalogs are URAT-1,
2MASS, and USNO-B1. This implies that PP is subject to the same
systematic and statistical offsets that are inherent to each of these
catalogs. For a discussion of these intrinsic offsets, please refer to
the corresponding catalog publications.

In order to quantify the astrometric uncertainty introduced by PP, the
positions of stars found in the image data used in Section
\ref{lbl:photometry_verification} are compared to the catalog
positions. Each field is registered using different catalogs and the
residuals in RA and Dec are
derived relative to those catalogs. Results are shown in Table
\ref{tab:astrometry}. Mean residuals are typically zero 
with standard deviations ${\leq}0.3$~arcsec. Star
positions from the images and the catalogs have been matched within
5~arcsec in order to minimize the number of false pairs. However, a
few matches still have residuals of a few arcsec, artificially
increasing the standard deviations. Generally, the URAT-1 catalog
provides more accurate positions than the other two
catalogs. This is most likely due to the fact that it is the most
recently published catalog relative to the date the image data have
been taken, minimizing the stars' proper motions.

The astrometric uncertainty introduced by PP is typically of the order
of 0.3~arcsec, depending on the utilized catalog, which allows for
accurate positional measurements of the targets. Note that positions
measured from images use Source Extractor's windowed centroids ({\tt
  XWIN\_WORLD}, {\tt YWIN\_WORLD}, see Source Extractor Manual) that
use a weighting scheme and provide positional precision similar to PSF
fitting routines.

\begin{table}
\scriptsize
\setlength{\tabcolsep}{5pt}
\begin{tabular}{|rcccc|}
\hline\\[-5pt]
 & \multicolumn{4}{c}{{\bf Positional Residuals} (arcsec)}\\
{\bf Field} &  & {\bf URAT-1} & {\bf 2MASS} & {\bf USNO-B1} \\
\hline\hline\\[-5pt]
RU149       &  RA  & 0.0$\pm$0.4 (503) & 0.0$\pm$0.4 (488)
& 0.0$\pm$0.5 (487)\\
       &  Dec  & 0.0$\pm$0.2 (503) & 0.0$\pm$0.3 (488)
& 0.0$\pm$0.4 (487)\\[+3pt]

PG0231+051       &  RA  & 0.0$\pm$0.1 (39) & 0.1$\pm$0.3 (51)
& 0.1$\pm$0.2 (53)\\
                 &  Dec  & 0.0$\pm$0.1 (39) & -0.1$\pm$0.3 (51)
& 0.0$\pm$0.4 (53)\\[+3pt]

PG1047+003       &  RA  & 0.0$\pm$0.1 (44) & 0.0$\pm$0.3 (43)
& -0.1$\pm$0.4 (42)\\
                 &  Dec  & 0.0$\pm$0.1 (44) & 0.0$\pm$0.2 (43)
& 0.1$\pm$0.7 (42)\\[+3pt]

PG1323-086       &  RA  & 0.0$\pm$0.1 (54) & 0.0$\pm$0.3 (53)
& 0.0$\pm$0.3 (53)\\
                 &  Dec & 0.0$\pm$0.2 (54) & 0.0$\pm$0.3 (53)
& 0.0$\pm$0.3 (53)\\[+3pt]

95-142           &  RA  & 0.0$\pm$0.2 (46) & 0.0$\pm$0.2 (47)
& 0.1$\pm$0.3 (43)\\
                 &  Dec & 0.0$\pm$0.1 (46) & 0.0$\pm$0.3 (47)
& -0.1$\pm$0.3 (43)\\[+3pt]

95-43           &  RA  & 0.0$\pm$0.1 (32) & 0.0$\pm$0.2 (31)
& 0.1$\pm$0.2 (32)\\
                &  Dec & 0.0$\pm$0.1 (32) & 0.0$\pm$0.3 (31)
& -0.1$\pm$0.3 (32)\\
\hline
\end{tabular}
\caption{Verification of the astrometric accuracy. Mean positional residuals and corresponding standard deviations based on comparing stars' positions from registered images and their catalogued positions; numbers in brackets are the number of stars used in the comparison. }
\label{tab:astrometry}
\end{table}

\subsection{Limitations of the Pipeline}
\label{lbl:limits}

PP has been designed to provide reliable photometric measurements for
the majority of imaging data taken with a large range of different
telescopes/instruments. This implies that PP will require manual 
  adjustments -- or fail entirely -- in a small fraction of possible
applications. The success of PP depends largely on the availability of
non-saturated background stars for the image registration and
photometric calibration. Hence, data taken with extraordinarily short
integration times, small fields of view, or under bad transparency
conditions might require manual interaction or fail entirely.

Furthermore, the pipeline relies on the availability of astrometric
and photometric catalogs to provide reliable and accurate
results. Current catalogs cover most of the Northern Hemisphere,
defining the application area of PP on the sky. The
astrometric and photometric accuracy provided by the pipeline is
mainly a function of the intrinsic accuracy of the catalogs used.  The
availability of high-accuracy catalogs in the near future (Pan-STARRS,
Gaia, LSST) will also improve the accuracy provided by PP.

PP was not designed to provide high-accuracy photometry such as might
be required for detecting exoplant transits. Instead, the main
objective of the pipeline is to provide reliable and calibrated
photometry on the ${\leq}$ 0.03~mag level for targets that are bright
enough.

\subsection{Future Developments}

The PP source code is maintained regularly, which includes the
implementation of newly available catalogs, as well as the setup of
additional telescope/instrument combinations. The pipeline is able to
support data from basically all professional and some
  high-level amateur telescopes that are able to provide the
necessary header information, including date and time of observation,
as well as telescope pointing information.  Future catalogs that will be
supported by PP include GAIA, Pan-STARRS, and LSST data for highly
improved astrometric and photometric calibration and a better sky
coverage. Additional transformations between different photometric
systems will be provided, as well.

Future releases of PP will also include additional support for
observations of a wide range of targets of interest. Calibrated
databases extracted from the FITS images can be matched with manually
created target catalogs or online resources (e.g., SIMBAD). The output
for each target can be submitted to large Virtual Observatory
databases for public access. Improved manual target selection based
directly on image coordinates will allow for the extraction of moving
targets even from data for which astrometric calibration is
impossible, e.g., through a lack of background stars, or significantly
trailed stars. This will enable observations of uncatalogued objects
and objects with large positional uncertainties, including satellites,
space debris, and not-yet confirmed near-Earth asteroids.

By utilizing SWARP\footnote{\url{http://www.astromatic.net/software/scamp}} \citep{Bertin2006},
PP will be able to stack images based on WCS coordinates -- also for
moving targets. The stacking greatly improves the signal-to-noise
ratio of the target.

Finally, PP will enable the identification and photometric measurement
of serendipitously observed asteroids in each field. The astrometric
accuracy of positions measured with the pipeline -- especially in
combination with catalogs as provided by GAIA -- will greatly improve
asteroid orbits at no additional cost. In addition to that, calibrated
photometric measurements will supplement pointed asteroid observations
and support efforts to find asteroid shapes and rotational periods
\citep[see, e.g.,][]{Durech2015}.


\vspace{2em}\par

PP was initially developed in the framework of the ``Mission
Accessible Near-Earth Object Survey'' (MANOS) and is supported by NASA
NEOO/SSO grants NNX14AN\-82G (MANOS; PI, N.\ Moskovitz, Lowell
Observatory) and NNX15AE90G (Rapid response observations of NEOs, PI:
D.\ E.\ Trilling, Northern Arizona University). The author would like
to thank N.\ Moskovitz for numerous discussions and testing the
pipeline in its different development stages, P.\ Massey for providing
the Discovery Channel Telescope LMI data of  standard fields, B.\
Skiff for useful discussions and analyzing the observations of
asteroid 2704 with MPO Canopus, D.\ Oszkiewicz for permission to use
this data set here, and D.\ E.\ Trilling for suggestions on the
  manuscript. The author would also like to thank E.\ Bertin and the
Astromatic team for providing their software to the public. 
  Finally, the author would like to thank an anonymous referee for
  useful suggestions and comments that improved this manuscript.

\section{Appendix: {\tt CALLHORIZONS} -- A Python Module to query JPL Horizons}
\label{lbl:callhorizons}

PP was originally designed for the analysis of asteroid
observations and therefore depends on ephemerides in order to
properly identify asteroids in the image data. These ephe\-merides are
obtained from the JPL Horizons system \citep{Giorgini1996}, providing
an accurate and reliable source of Solar System ephemerides. Horizons
can be manually queried through a web
interface\footnote{\url{http://ssd.jpl.nasa.gov/horizons.cgi}} or a
telnet interface.

In order to be able to query ephemerides for a large number of
asteroids and dates into a Python environment, I created
CALLHORIZONS\footnote{\url{https://github.com/mommermi/callhorizons}}. This
Python module allows to query Horizons using its web
interface in order to obtain ephemerides and orbital elements for
given dates or date ranges. All objects in the Horizons database can
be queried, including planets, asteroids, comets, and spacecraft. The
query results are provided as NumPy arrays, providing a large degree
of flexibility in the analysis. Also, CALLHORIZONS provides a direct
interface to the PyEphem
module\footnote{\url{http://rhodesmill.org/pyephem/}}: orbital
elements queried by CALLHORIZONS can be directly turned into PyEphem
objects, enabling the user to calculate ephemerides locally.

\bibliography{mybibfile}

\end{document}